\documentclass[10pt,reprint]{iopart}

\usepackage[pdftex]{graphicx}
\usepackage{bm}
\usepackage{amssymb}

\usepackage{lineno}

\newcommand{\F}{\mathcal{F}}

\newcommand{\s}{\mathcal{S}}

\newcommand{\B}{\mathcal{B}}
\renewcommand{\H}{\mathcal{H}}
\newcommand{\R}{\bf{R}}

\newcommand{\Z}{\bf{Z}}

\newcommand{\E}{\mathbb{E}}

\newcommand{\re}{\mathrm{Re}}
\newcommand{\arrow}{\rightarrow}

\newcommand{\as}{\mathrm{-a.s.}}
\newcommand{\dPdQphi}{\frac{dP}{d(\varphi Q)}}

\newcommand{\dPphinvdQ}{\frac{d(\varphi^{-1}P)}{dQ}}
\newcommand{\phinv}{\varphi^{-1}}
\newcommand{\Pphinv}{\varphi^{-1} P}
\newcommand{\Qphi}{\varphi Q}
\renewcommand{\P}{P_{[-t,t]}}

\begin{document}

\title{The Measure-theoretic Identity Underlying Transient Fluctuation Theorems}

\author{Benjamin Hertz Shargel}
\address{Department of Mathematics, UCLA, Los Angeles, CA, 90095-1766\\}
\ead{shargel@math.ucla.edu}


\begin{abstract}
We prove a measure-theoretic identity that underlies all transient fluctuation theorems (TFTs) for entropy production and dissipated work in inhomogeneous deterministic and stochastic processes, including those of Evans and Searles, Crooks, and Seifert.  The identity is used to deduce a tautological physical interpretation of TFTs in terms of the arrow of time, and its generality reveals that the self-inverse nature of the various trajectory and process transformations historically relied upon to prove TFTs, while necessary for these theorems from a physical standpoint, is not necessary from a mathematical one.  The moment generating functions of thermodynamic variables appearing in the identity are shown to converge in general only in a vertical strip in the complex plane, with the consequence that a TFT that holds over arbitrary timescales may fail to give rise to an asymptotic fluctuation theorem for any possible speed of the corresponding large deviation principle.  The case of strongly biased birth-death chains is presented to illustrate this phenomenon.  We also discuss insights obtained from our measure-theoretic formalism into the results of Saha et. al. on the breakdown of TFTs for driven Brownian particles. 
\end{abstract}
\pacs{02.50.Ey, 05.70.Ln}

\maketitle

\flushleft

\bigskip
\bigskip


\section{Introduction}

\quad Fluctuation theorems have drawn a significant amount of research attention since their discovery over fifteen years ago \cite{Evans1993,Gallavotti} due to their apparent connection to irreversibility in nonequilibrium processes.  Whereas the second law states that the expected value of thermodynamic quantities such as entropy production and dissipated work must be nonnegative, these theorems reveal a symmetry in the actual probability distributions of these quantities, of which the nonnegative expectation is merely one consequence.  Transient fluctuation theorems (TFTs), which provide concrete probabilities of second law violations in systems observed at finite length and time scales, have even been credited \cite{Sevick} as the resolution of Loschmidt's paradox: how microscopically reversible dynamics can give rise to macroscopically irreversible phenomena.   

\quad In this paper we show that, beyond their known connection to the second law, TFTs in fact possess a tautological physical interpretation in terms of the arrow of time.  This result rests on the derivation of a measure-theoretic identity that underlies all TFTs for entropy production and dissipated work, but not, notably, heat dissipation - the time-extensive current part of entropy production.  In particular, we will see that entropy production and dissipated work satisfy the identity (and, hence, a TFT) solely because of their representation as logarithmic Radon-Nikodym derivatives, revealing that the self-inverse nature of the protocol \cite{Crooks,Jarzynski}, trajectory \cite{Maes2003}, driving field \cite{Maes2006} and process adjoint \cite{Esposito,Harris} transformations that has previously been relied upon to prove TFTs, while necessary for these theorems from a physical standpoint, is not necessary from a mathematical one.  The identity underlying TFTs, in fact, permits general noninvolutive process and trajectory transformations that are unrelated to the irreversibility of the underlying process. 

\quad As an example of the generality we claim for TFTs, consider a physical system represented in reduced coordinates by a continuous time Markov chain on the finite state space $\{1,2,\dots,N\}$, satisfying local detailed balance and driven by time-dependent, positive transition rates $k_{ij}:[-t,t] \rightarrow (0,\infty)$.  Suppose further that the system is initially prepared at time $-t$ in an equilibrium distribution that satisfies strict detailed balance with respect to the rates $k_{ij}(-t)$.  It is a well-known result \cite{Crooks,Harris} that the TFT relation $f(x)/f^B(-x) = e^x$ then holds between the probability density $f$ of the dissipated work of this process (work done on the system that is not stored as free energy but released as heat) and the density $f^B$ of the dissipated work of the corresponding backward process, in which the sample paths and transition rates have been time-reversed.  Surprisingly, however, the exact same relation holds when the transition rates of the backward process are replaced by an {\it arbitrary} driving protocol $k_{ij}':[-t,t] \rightarrow (0,\infty)$ and the path-reversal transformation replaced by one which dices up a path according to an arbitrary finite partition and rearranges it, preserving right-continuity.  While the first TFT appears related to the irreversibility of the original process, the second one clearly is not. 

\quad Several advances have already been made in the pursuit of a generalized, or universal, TFT.  In a sweeping series of papers, Maes and collaborators proved a moment generating function (MGF) symmetry for the entropy production of general classes of stochastically \cite{Maes1999,Maes2000,Maes2003,Maes2006,Maes2008} and deterministically \cite{Maes2000,Maes2004} modeled homogeneous processes, in which entropy production was identified as the source term for time-reversal breaking in the process' action functional and found to equal the logarithmic Radon-Nikodym derivative of the process' path measure with respect to itself, composed with a path-reversal transformation.  Ge and Jiang later rigorously proved the corresponding distributional form of this symmetry \cite{GeJiang}, which is the one most often found in applications and experimental studies of TFTs (see e.g., Refs. \cite{Evans1993,Kim,Lahiri,Wang}).  The distributional form of the symmetry was generalized in a nonrigorous fashion by Crooks \cite{Crooks} to inhomogeneous stochastic processes satisfying local detailed balance \cite{KatzLebowitzSpohn,Lebowitz} via the introduction of a protocol-reversed process, and was generalized by Jarzynski \cite{Jarzynski} to the case of inhomogeneous Hamiltonian systems connected to multiple thermal reservoirs.  In their review paper, Harris and Sch\"utz \cite{Harris} generalized both forms of the TFT to the case of inhomogeneous Markov chains using a general functional on the Markov chain path space that consists of a current and boundary part \cite{Seifert}, which is able to represent various thermodynamic quantities depending on the choice of the latter.  This flexibility allowed them to recreate a Markov chain version of many of the existing TFTs in the literature.

\quad The TFT identity we prove in this paper builds off and extends all of these results and subsumes all integral and transient fluctuation theorems proven to date, including those 
of Evans and Searles \cite{Evans2002,Searles1999} and Seifert \cite{Seifert}.  Like existing TFTs, it can be expressed in both distributional and MGF form.  It holds for the entropy production and dissipated work of all inhomogeneous deterministic and stochastic processes satisfying local detailed balance, including processes for which the distributions of these thermodynamic quantities are neither singular nor continuous with respect to Lebesgue measure.  From the measure-theoretic formalism we use to prove the identity, however, it is clear that it does not in general hold for the heat dissipated by a process, a fact observed by van Zon and Cohen \cite{vanZon2004} in the context of driven Brownian particles.  Similar reasoning yields intuition into some of the more recent results of Saha et. al. \cite{Saha}.

\quad An additional result that follows from our analysis is that the MGFs of thermodynamic variables appearing in TFTs, presumed by previous studies to exist everywhere, in fact converge in general only on the vertical strip $-1 \leq \re(\lambda) \leq 0$ in the complex plane.  As we discuss in section 4, the failure of an MGF to converge in a neighborhood of the origin rules out the possibility of a large deviation principle for the associated variable, regardless of the speed, or time-scaling, used in its formulation.  The implication is that thermodynamic quantities may satisfy a TFT over arbitrary timescales, and yet not satisfy an asymptotic fluctuation theorem (AFT), complementing our previous results on the breakdown of AFTs for even continuous and bounded driving protocols \cite{Shargel}.  Note that this type of AFT breakdown is different from those considered by van Zon and Cohen \cite{vanZon2004}, Baiesi et. al \cite{Baiesi} and Rakos and Harris \cite{Rakos}, in which the time-averaged heat dissipation of a process satisfies a large deviation principle with linear speed, but whose rate function exhibits distinct, "extended" fluctuation symmetries over different regions of its domain. 

\quad The rest of the paper is organized as follows.  In section 2 we present our abstract formulation of the TFT identity, beginning with the definition of logarithmic Radon-Nikodym derivatives $S_P$ and $S_Q$, which take the place of Harris and Sch\"utz's generalized path functionals. Both forms of the identity as well as a mutual implication are then proved for these quantities, paying careful attention to the domain of convergence of the MGFs.  In section 3 we identify $S_P$ (resp. $S_Q$) as either the entropy production or dissipated work of the forward (resp. backward) version of a process, depending on the choice of its boundary term, confirming that these thermodynamic quantities satisfy the TFT identity solely by virtue of their representation as a logarithmic derivative.  This is followed by a discussion of the tautological physical interpretation of TFTs for homogeneous and inhomogeneous processes.  In section 4 we switch gears to consider the breakdown of asymptotic fluctuation theorems due to the divergence of MGFs, illustrating with the example of strongly biased birth-death chains.  Section 5 concludes with summary remarks.

\section{Abstract definitions and results}

\quad Consider a pair of (nearly) arbitrary probability spaces $(\Omega,\F,P)$ and $(S,\s,Q)$.  While the theorems we will prove are entirely general, making no assumptions about the structure of $(\Omega,\F)$ or $(S,\s)$, we will always have in mind the case in which $(\Omega,\F)=(S,\s)$ and elements $\omega \in \Omega$ are spacetime trajectories of some deterministic or stochastic process.  In this scenario, $P$ will play the role of the forward process measure and $Q$ the protocol-reversed process measure, in which any inhomogeneities driving the forward process have been time-reversed.  We assume the existence of a bimeasurable transformation $\varphi:\Omega \arrow S$ between the two sample spaces (i.e., $\varphi$ and $\varphi^{-1}$ are both measurable functions), whose action on $\omega \in \Omega$ will be denoted $\varphi \omega$ and is defined on sets $A \subset \Omega$ by $\varphi A = \{\varphi \omega: \omega \in A\}$ and measures $\mu$ by $\varphi \mu(A) = \mu(\varphi(A))$.  Our central objects of interest will be the random variables $S_P = \log dP/d(\varphi Q)$ and $S_Q = \log dQ/d(\varphi^{-1} P)$, which generalize both entropy production and dissipated work under the forward and backward process, respectively.  The precise connection between these variables and the physical quantities they represent depends on the initial distributions we assign to the forward and backward process measures, and will be addressed in section 3.  

\quad Recall that we say a probability measure $\mu$ is absolutely continuous with respect to another probability measure $\nu$, and write $\mu \ll \nu$, if $\nu(A) = 0$ implies $\mu(A) = 0$ for all events $A$.  We say that two such measures are equivalent, and write $\mu \sim \nu$, when the two are mutually absolutely continuous.  When $\mu$ and $\nu$ are path measures, as is the case in this paper, this means that the two measures put positive probability on the same set of paths.  We begin this section by introducing a lemma that demonstrates how composition with bimeasurable transformations preserves the absolute continuity of measures and affects the corresponding Radon-Nikodym derivative, which plays the role of a Jacobian between the two measures.  The lemma will be an important symbolic tool in the theorems to follow because, loosely speaking, it allows us to replace a $\varphi$ in front of $Q$ with a $\varphi^{-1}$ in front of $P$, and vice-versa.  Note that statement (c) is a generalization of Lemma 2.1 in \cite{GeJiang}.

\bigskip

{\bf Lemma 2.1.} {\it Given the definitions above, the following three statements hold.}

\begin{eqnarray}
\label{LEMMA_A}
& \mathrm{(a)} \, P \ll \Qphi \Longrightarrow \Pphinv \ll Q \\
\vspace{5mm}
\label{LEMMA_B}
& \mathrm{(b)} \, P \sim \Qphi \iff \Pphinv \sim Q \\
\vspace{5mm}
\label{LEMMA_C}
& \mathrm{(c)} \, P \ll \Qphi \Longrightarrow \dPdQphi(\phinv \omega) = \dPphinvdQ(\omega), 
\,\, Q \mathrm{-} a.s.
\end{eqnarray}

{\bf Proof.} Let $A \in \s$ in what follows.  If $P \ll \Qphi$, then $Q(A) = \Qphi(\phinv A) = 0$ $\Longrightarrow P(\phinv A) = 0$ $\Longrightarrow \Pphinv(A) = 0$. This proves (a).  Now taking $\Qphi \ll P$,  $\Pphinv(A) = P(\phinv A) = 0$ $\Longrightarrow Q(A) = \Qphi(\phinv A) = 0$, implying $Q \ll \Pphinv$, and hence the forward implication of b).  Similar arguments yield the reverse implication.   Finally, assuming $P \ll \Qphi$, since $dP/d(\Qphi(\phinv \, \cdot))$ is $Q$-measurable,  

\begin{equation*} 
\int_A \dPdQphi(\phinv \omega)dQ(\omega) 
= \int_{\phinv A} \dPdQphi(\omega')dQ(\varphi \omega')
= \int_{\phinv A} dP(\omega'),
\end{equation*}

which guarantees existence of the left-hand integral.  We further have

\begin{equation*}
P(\phinv A) = \Pphinv(A) = \int_A \dPphinvdQ(\omega)dQ(\omega),
\end{equation*}

where the second equality follows from (a). By uniqueness of the Radon-Nikodym derivative up to sets of measure zero, the result follows. 

\bigskip

\quad We now prove our first main result, the MGF form of our TFT identity, which generalizes the MGF symmetries of Maes et. al. and Harris and Sch\"utz.  Note the mutual independence of the two measures and $\varphi$: neither the processes themselves nor the transformation between them need be related beyond a mutual absolute continuity assumption for the symmetry to hold.  This is in contrast to existing results for inhomogeneous processes, for which the backward process $Q$ is defined directly in terms of the original process $P$ and path-reversal is often not replaceable by a more general transformation.  The fact that $\varphi$ is merely assumed to be bimeasurable is itself a new result, in particular, since all existing proofs of TFTs rely on the involutive, or self-inverse, nature of time reversal.   

\bigskip

{\bf Theorem 2.2 (MGF symmetry).}  {\it Given probability spaces $(\Omega,\F,P)$ and $(S,\s,Q)$ with a bimeasurable transformation $\varphi:\Omega \arrow \s$ between them such that $P \sim \Qphi$},

\begin{equation} 
\E_P \bigl( e^{\lambda S_P} \bigl) = \E_Q \bigl( e^{-(1+\lambda) S_Q} \bigl)
\quad \mathrm{for} \,\, -1 \leq \re(\lambda) \leq 0. 
\label{GFT}
\end{equation}

{\bf Proof.} \, We first show existence of the generating functions.  Expanding out their definitions,

\begin{eqnarray} 
\bigl| \E_P \bigl( e^{\lambda S_P} \bigl) \bigl| \, 
& = \biggl| \int_\Omega \biggl( \dPdQphi(\omega) \biggl)^{\lambda} dP(\omega) \biggl| \nonumber \\
& \leq \int_\Omega \biggl( \dPdQphi(\omega) \biggl)^{\re(\lambda)} dP(\omega)
\label{EP}
\end{eqnarray}

and

\begin{eqnarray} 
\bigl| \E_Q \bigl( e^{-(1+\lambda) S_Q} \bigl) \bigl| \, 
& = \biggl| \int_S \biggl( \dPphinvdQ(\omega) \biggl)^{1+\lambda} dQ(\omega) \biggl| 
\nonumber \\
& \leq \int_S \biggl( \dPphinvdQ(\omega) \biggl)^{1+\re(\lambda)} dQ(\omega)
\label{EQ}
\end{eqnarray}

by Jensen's inequality, where the derivative in (\ref{EQ}) is justified by (\ref{LEMMA_A}).  Recall that by finiteness of the measures $P$ and $Q$, the spaces $L^p(P)$ and $L^p(Q)$ have the property that $L^{p_2} \subset L^{p_1}$ for $0 \leq p_1 \leq p_2$.  Noting that $d(\varphi Q)/dP \in L^1(P)$, we therefore have that $d(\varphi Q)/dP \in L^p(P)$ for $p \in [0,1]$, or, by equivalence of $P$ and $\varphi Q$, $dP/d(\varphi Q) \in L^{\re(\lambda)}(P)$ for $\re(\lambda) \in [-1,0]$.  Similarly, $d(\phinv P)/dQ \in L^p(Q)$ for $p \in [0,1]$, and so $d(\phinv P)/dQ \in L^{1+\re(\lambda)}(Q)$ for $\re(\lambda) \in [-1,0]$.  This gives convergence of both MGFs for $\re(\lambda) \in [-1,0]$.  For $\lambda$ in this domain,

\begin{equation}
\label{MGF_EQUALITY}
\eqalign{
\E_P \bigl( e^{\lambda S_P} \bigl) 
& = \int_\Omega \biggl( \dPdQphi(\omega) \biggl)^{\lambda} dP(\omega) \\
& = \int_\Omega \biggl( \dPdQphi(\omega) \biggl)^{\lambda} \dPdQphi(\omega) 
	d(\Qphi)(\omega) \\
& = \int_\Omega \biggl( \dPdQphi(\omega) \biggl)^{1+\lambda} dQ(\varphi \omega) \\
& = \int_S \biggl( \dPdQphi(\phinv \omega') \biggl)^{1+\lambda} dQ(\omega') \\
& = \int_S \biggl( \dPphinvdQ(\omega') \biggl)^{1+\lambda} dQ(\omega') \\
& = \E_Q \bigl( e^{-(1+\lambda) S_Q} \bigl),
}
\end{equation}

where the final two equalities follow from (\ref{LEMMA_C}) and then (\ref{LEMMA_B}). 

\bigskip

\quad The distributional form of the TFT identity is the one most often cited in applications and experimental studies of fluctuation theorems.  Existing versions of it require the distributions of thermodynamic variables (our $S_P$ and $S_Q$) to be singular or continuous with respect to Lebesgue measure, but in our generalization below we relax this assumption to allow for more general processes whose distributions of these quantities may be more complicated.  To state it, we define $dF$ and $dG$ to be the Lebesgue-Stieltjes measures corresponding to the distribution functions $F(x) = P(S_P \leq x)$ and $G(x) = Q(-S_Q \leq x)$, respectively.

\bigskip

{\bf Theorem 2.3 (Distributional symmetry).}  {\it Under the conditions of Theorem 2.2},

\begin{equation}
dF \sim dG \quad \mathrm{ and } \quad \frac{dF}{dG}(x) = e^x
\label{GTFT1}
\end{equation}

{\bf Proof.}  We first note that we may represent these measures as $dF = dP \circ S_P^{\, -1}$ and $dG = dQ \circ (-S_Q)^{-1}$.  Given a Borel set $A \subset \R$, invoking the lemmas as before yields

\begin{eqnarray*}
\int_A dF(x) 
& = P(S_P \in A) \\
& = P \{ \omega \in \Omega : \log \dPdQphi(\omega) \in A \} \\
& = \Pphinv \{ \omega' \in S : \log \dPdQphi(\phinv \omega') \in A \} \\
& = \Pphinv \{ \omega' \in S : \log \dPphinvdQ(\omega') \in A \}, \\
\end{eqnarray*}

where in the final equality we have used $\phinv P \sim Q$ to justify that (\ref{LEMMA_C}) holds $(\phinv P)\as$  By (\ref{LEMMA_B}), the last expression equals 

\begin{eqnarray*} \fl
\Pphinv (-S_Q \in A)
& = \int_{-S_Q \in A} \dPphinvdQ(\omega) dQ(\omega)
  = \int_{-S_Q \in A} e^{-S_Q(\omega)} dQ(\omega) \\
& = \int_A e^{x} dQ((-S_Q)^{-1}(x))
  = \int_A e^{x} dG(x).
\end{eqnarray*}

We therefore conclude $dF \ll dG$ and $dF/dG = e^x$.  By finiteness and strict positivity of the Radon-Nikodym derivative, however, it immediately follows that $dG \ll dF$, and so $dG \sim dF$.  \qquad 

\bigskip

{\bf Corollary 2.4.}  {\it When $S_P$ and $S_Q$ both have continuous or discrete distributions under $P$ and $Q$, respectively, then under the conditions of Theorem 1},

\begin{equation}
P(S_P = x) = e^x \, Q(S_Q = -x).
\label{GTFT2}
\end{equation}

{\it Here $P(S_P = x)$ and $Q(S_Q = -x)$ denote densities in the case of continuous distributions and probability mass functions in the case of discrete distributions.}  

\bigskip

\quad It was argued informally in Ref. \cite{Harris} that the MGF form of the identity implies the distributional form.  We rigorously prove this statement here, as well as its converse, when the real part of $\lambda$ is restricted to $[-1,0]$.  This is not an {\it a priori} obvious fact, since classical statements of the uniqueness of MGFs require them to be defined in a neighborhood of the origin.  We will see in section 4 that this domain of convergence cannot, in general, be extended.

\bigskip

{\bf Theorem 2.5 (Equivalence of the MGF and distributional symmetries).}

\begin{equation*}
dF \sim dG \quad \mathrm{ and } \quad \frac{dF}{dG}(x) = e^x
\end{equation*}

{\it holds if and only if}

\begin{equation*} 
\E_P \bigl( e^{\lambda S_P} \bigl) = \E_Q \bigl( e^{-(1+\lambda) S_Q} \bigl)
\quad \mathrm{for} \,\, -1 \leq \re(\lambda) \leq 0. 
\end{equation*}
  
{\it This statement is not true if the domain of $\lambda$ is extended.}  
  
\bigskip

{\bf Proof.}  Beginning with the reverse implication, we saw in the proof of (\ref{GFT}) that $\E_P \bigl( e^{\lambda S_P} \bigl)$  and $\E_Q \bigl( e^{-(1+\lambda) S_Q} \bigl)$ exist for $-1 \leq \re(\lambda) \leq 0$.  For such values of $\lambda$,

\begin{equation}
\eqalign{
	\int_{\R} e^{\lambda x} dF(x)
		& = \E_P \bigl( e^{\lambda S_P} \bigl) \\
		& = \E_Q \bigl( e^{-(1 + \lambda) S_Q} \bigl) \\
		& = \int_{\R} e^{(1+\lambda) x} dG(x) \\
		&  = \int_{\R} e^{\lambda x} e^x dG(x)}
\label{EQUALITIES}
\end{equation}

We would like to identify the measures $dF$ and $e^x dG$, as this equality holds for a continuum of $\lambda$ values.  To do so, we begin by defining $\lambda' = \lambda + 1/2$, $d\mu = e^{-x/2} dF$ and $d\nu = e^{x/2} dG$.  Then, by (\ref{EQUALITIES}), 
$f(\lambda') = \int_{\R} e^{\lambda' x} d\mu(x)$ and $g(\lambda') = \int_{\R} e^{\lambda' x} d\nu(x)$ exist and are equal in the closed strip $-1/2 \leq \re(\lambda') \leq 1/2$.  This domain contains the imaginary axis, on which $f(-\lambda')$ and $g(-\lambda')$ are the characteristic functions of $\mu$ and $\nu$, respectively.  As characteristic functions uniquely determine their measures \cite{Durrett}, we have $\mu = \nu$.  The strict positivity of the exponentials in the definition of $\mu$ and $\nu$ then implies that $dF \sim dG$, with $dF/dG = e^x$.  The forward implication of the theorem is trivial by rearranging the equalities in (\ref{EQUALITIES}), given the existence of the generating functions for $\re(\lambda) \in [-1,0]$.   

\quad For the proof of why the mutual implication in the theorem does not remain valid if the domain of $\lambda$ is extended, we refer the reader to section 4.2, in which we discuss the example of a strongly biased birth-death chain run from time $0$ to $t$.  The transition rates of the process are defined such that the forward measure $P$ and backward measure $\varphi Q$ of the process are equivalent, and so by Theorem 2.2, $dF \sim dG$ with $dF/dG = e^x$.  However, the MGF of $S_P$ is shown to diverge for $\re(\lambda) > 0$, which implies that, by equality (\ref{MGF_EQUALITY}), the MGF of $S_Q$ diverges for $\re(\lambda) < -1$.


\section{Application to deterministic and stochastic processes}  

\subsection{Forward and backward processes}

\quad Having proved general results for the abstract quantities $S_P$ and $S_Q$, our goal is now to connect them directly to the entropy production and dissipated work of deterministic and stochastic processes, in order to confirm that our TFT identity does indeed subsume the most general TFTs in the physics literature.  To begin with, we must identify the building blocks of these quantities, the measures $P$ and $Q$ and the transformation $\varphi$, as well as the measurable spaces $(\Omega, \F)$ and $(S, \s)$ on which they are defined.  

\quad Due to the generality of the class of processes we wish to consider, our definitions will be open-ended.  $(\Omega, \F) = (S, \s)$ we take to be a measurable space of functions $\omega$ from $[-t,t]$ to a state space $(\Xi,\B)$, in which the path space $\Omega$ may be, for example, $C([-t,t],\Xi)$, $D([-t,t],\Xi)$ (the Skorokhod space of right-continuous paths with left limits) or a Riemannian manifold, depending on the underlying dynamics.  The measure $P \equiv P_{[-t,t]}$ governs the forward (original) process, which in general depends upon a spacetime-dependent protocol $\lambda(x,s)$ ($x \in \Xi$ and $-t \leq s \leq t$), and $Q \equiv \P^-$ then governs the process in which the protocol has been time-reversed.  $\varphi \equiv r$ we take to be an appropriately defined path-reversal involution (i.e., $r^{-1} = r$), which may, for example, need to preserve right-continuity of paths or reverse the momentum coordinates of a Hamiltonian state space.  The minimal requirement that this transformation be bijective between the supports of the measures $\P$ and $\P^-$ is at times called "dynamic reversibility" \cite{Maes2003} and "microscopic reversibility" \cite{Sevick}.

\quad The composition $\P^B \equiv r \P^-$ we call the backward path measure, and may be interpreted as follows.  For a subset $A \subset \Omega$, as $\P(A)$ is the probability of one of the spacetime curves in $A$ being realized in a universe in which time runs forward from $-t$ to $t$, $\P^B(A)$ is the probability of one of those curves being realized in a universe in which time runs {\it backward} from $t$ to $-t$.  (For this reason we refer to $\omega_t$ as the initial state of the backward process, and $\omega_{-t}$ the final state.)  This is in contrast to the case of time-symmetric protocols (including, particularly, time-independent ones), in which $\P^B(A) = r \P(A)$ becomes simply the probability of observing the reverse of some $\omega \in A$ as time moves in the usual forward direction.  Equality between $\P$ and $r\P$ in the homogeneous case is how macroscopic reversibility is usually defined.  Note that microscopic reversibility amounts to equivalence between the measures $\P$ and $\P^B$.  

\quad It has been well-established that when local detailed balance is satisfied, the entropy production and dissipated work of inhomogeneous Markov chains \cite{Harris} and the entropy production of homogeneous diffusions \cite{JiangBook, Maes2000, Qian} equal the logarithmic Radon-Nikodym derivative of their forward path measure with respect to their backward path measure, with suitably chosen initial distributions.  Our goal here is to argue that this representation is universal and well-defined for the dissipated work and entropy production of general stochastic and deterministic processes satisfying local detailed balance.  

\quad What distinguishes entropy production, dissipated work and dissipated heat in deterministic and stochastic processes are the initial distributions for the forward and backward path measures.  This was first demonstrated in the case of Langevin processes \cite{Seifert}, and then later for Markov chains \cite{Harris}.  In order to define these quantities for the more general classes of processes to follow, we adopt the following terminology.  Let $\mu(\cdot,s) = \P(X_s \in \cdot)$ denote the law of the forward process and $\mu^B(\cdot,s) = \P^B(X_s \in \cdot)$ the law of the backward process for $s \in [-t,t]$, where $X_s(\omega) = \omega_s$ is the coordinate projection from $\Omega$ onto $\Xi$.  These measures must both be either continuous densities or discrete distributions $\forall \, s$ in order for the system entropy $\log \mu(X_s(\omega),s)$ to be a well-defined quantity, and may therefore be obtained by solving Kolmogorov forward equations.  Microscopic reversibility further implies that they have the same support, a condition sometime called ergodic consistency \cite{Evans2002,Sevick}.  For processes possessing a Hamiltonian $H(x,s) = H_{\lambda(x,s)}(x)$ on their state space, let $\mu^*(dx,s) =$ $\exp[- \beta H(x,s)]dx/Z(s)$ denote the Gibbs measure corresponding to equilibrium when the protocol is held fixed in time at $\lambda(\cdot,s)$, where $\beta$ denotes a possibly non-physical inverse temperature of a connected thermal reservoir and $Z(s) = Z_{\lambda(\cdot,s)}$ the partition function corresponding to $H(\cdot,s)$.   

\subsection{Entropy production}

\quad The first claim we make is that $S_P = \log d\P/d\P^B$ equals the entropy production $S(-t,t)$ of the forward process accumulated from time $-t$ to $t$ when local detailed balance is satisfied and we impose the boundary condition 

\begin{equation}
\label{BC1}
\mathrm{(BC1)} \quad \mu^B(\cdot,t) = \mu(\cdot,t),
\end{equation}

that is, when the forward and backward laws agree at the final time $t$.  Indeed, this was first argued in Ref. \cite{Crooks} for general stochastic processes, proved later in the context of inhomogeneous Markov chains \cite{Harris}, and is consistent with later measure-theoretic \cite{JiangBook,Maes2000,Qian} and Onsager-Machlup \cite{Maes2003} analyses of homogeneous diffusions. 

\quad The argument for general Markov processes goes as follows.  Coarse-graining the state space $\Xi$ of the process into discrete states $i \in \Z$, local detailed balance implies that for all times $s \in [-t,t]$, 

\begin{equation}
\label{INCR_HEAT}
\frac{p(i,j,s)}{p(j,i,s)} = e^{\beta \Delta Q(s)},
\end{equation}

where $p(i,j,s)$ is the rate of transitioning from $i$ to $j$ at time $s$ and $- \Delta Q(s)$ is the heat which must be imported from a reservoir at inverse temperature $\beta$ for the system to make this transition.  For systems with a Hamiltonian, $- \Delta Q(s) =$ $H(j,s) - H(i,s) + \Delta W(s)$, where $\Delta W(s)$ is the work that must be done against a nonconservative external force, such as an electric field in the presence of periodic boundary conditions \cite{KatzLebowitzSpohn} to make the transition happen.  In the absence of nonconservative forces, $\Delta W(s)$ vanishes and (\ref{INCR_HEAT}) reduces to strict detailed balance with respect to $H(\cdot,s)$.  For systems without a Hamiltonian, for whom a thermodynamic description is simply an idealization, $\Delta Q(s)$ has no meaning by itself, but $\beta \Delta Q(s)$ represents the entropy lost by the system as a result of the transition.  This is analogous to phase space contraction in dissipative dynamical systems, discussed below.

\quad Multiplying terms of the form (\ref{INCR_HEAT}) for every transition made from $-t$ to $t$ along a trajectory $\omega$ (in addition to exponential holding time factors on top and bottom that cancel), we find that the net heat $Q(-t,t)$ {\it exported} over the entire trajectory satisfies

\begin{equation}
\label{HEAT}
Q(-t,t) = \beta^{-1} \log \frac{\P(\omega|X_{-t}(\omega))}{\P^B(\omega|X_t(\omega))},
\end{equation}

where we have conditioned the forward and backward measures on the initial and final state of the trajectory, respectively.   Adding the net change in entropy of the system from $-t$ to $t$

\begin{equation*}
\Delta S = \log \frac{\mu(X_{-t}(\omega),-t)}{\mu(X_t(\omega),t)}
\end{equation*}

and using BC1, we find that the total entropy production

\begin{equation*}
S(-t,t) = \beta Q(-t,t) + \Delta S = \log \frac{d\P}{d\P^B}.
\end{equation*}

\quad This confirms that our identity subsumes the TFTs for entropy production of Crooks \cite{Crooks} and Seifert \cite{Seifert}.  Note that the above argument does not depend essentially on the coarse-graining of the state space $\Xi$, used here for simplicity, and that continuous system movements are fine so long as the appropriate local detailed balance condition is satisfied.

\quad We now consider the case of a nonconservative deterministic process, driven by a time-dependent potential and/or external dissipative field, and possibly in contact with a Nos\'e-Hoover or other deterministic thermostat, modulating fluctuations in momentum (see \cite{Sevick} for a thorough discussion).  The path space is taken to be a smooth submanifold of $\Xi^{[-t,t]}$ consisting of paths satisfying the thermostatted equations of motion, and the measures $\P$ and $\P^B$ are induced directly from the initial distributions $\mu(\cdot,-t)$ and $\mu^B(\cdot,t)$, which are presumed to possess densities $f$ and $f^B$, respectively, with respect to the underlying Lebesgue measure on the manifold.  The Evans-Searles dissipation function $\Omega(-t,t)$ \cite{Evans2002} of the forward process, while defined originally only for homogeneous processes, becomes in our inhomogeneous setup

\begin{equation*} \fl
\Omega(-t,t) = \log \frac{d\P}{d\P^B}
	= \log \frac{d\mu(X_{-t},-t)}{d\mu^B(X_t,t)}
	= \log \frac{f(X_{-t},-t)}{f^B(X_t,t)} + \int_{-t}^t \Lambda(X_s,s)ds,
\end{equation*}

where $\Lambda(X_s,s) = - \partial/\partial X_s \cdot \dot{X}_s$ is the phase space compression factor.  The first term on the RHS is the change in system entropy $\Delta S$, and the second equals the net phase space contraction over the trajectory, which, for many choices of thermostat \cite{Sevick}, equals the outward entropy flux $\beta Q(-t,t)$.  By the second law, we may therefore again conclude that $S_P = \Omega_P = S(-t,t)$.  

\quad The identification of $S_P$ with entropy production makes sense in two important respects.  First, in light of (\ref{BC1}), for each sample path $\omega$, $S_P(\omega)$ represents the log-likelihood of observing that path as time runs forward from $-t$ to $t$ rather than its {\it reverse} as time runs {\it backward} from $t$ to $-t$, vanishing precisely when $\P = \P^B$ - that is, when there is no way to tell from observing the process whether time is moving forward or backward.  Note that $S_P$ vanishes if and only if Feng and Crooks' time asymmetry $A =$ $\frac{1}{2} \H(\P,\frac{1}{2}(\P + \P^B)) +$ $\frac{1}{2} \H(\P,\frac{1}{2}(\P + \P^B))$ \cite{Feng}, a proposed measure of time's arrow, does.  Second, the expectation $\E_P(S_P) = \H(\P,\P^B)$ is nonnegative, consistent with the second law of thermodynamics, and, as observed by Maes \cite{Maes2003}, is equal up to coarse-graining corrections to the Gibbs entropy production of the process.  

\quad BC1 has an experimental connection to entropy production as well.  Imagine that we observe a large number of realizations of a process and wish to determine whether the  process is evolving forward or backward through time (i.e., whether the trajectories are being sampled from the forward or backward path measure).  For a simple homogeneous example, suppose we observe the temperature profile of a slab of material evolve from a Gaussian of width $\sigma_1$ at time $-t$ to one of width $\sigma_2 < \sigma_1$ at time $t$.  This is consistent with standard diffusion, but under the backward, not forward, path measure governing the microscopic dynamics.  To have generated the observed trajectories, the "initial" distribution $\mu^B(\cdot,t)$ of the backward path measure and the final distribution of the forward path measures must both equal the empirical one generated by the states of the trajectories at time $t$, precisely the boundary condition BC1.

\quad Having firmly established $S_P$ as the entropy production of the forward process under BC1, the question remains whether its counterpart $S_{P^-}$ ($S_Q$ in section 2) equals entropy production under the backward process.  While our TFT identity is valid for $S_{P^-}$ generally, only when this latter identification holds does our identity become a proper TFT for entropy production.  In fact, it was shown first by Crooks \cite{Crooks} and then at greater length by Harris and Sch\"utz \cite{Harris} that the path functionals $S_P$ and $S_{P^-}$ represent the same physical quantity under the forward and backward path measures, respectively, only when the initial distributions $\mu(\cdot,-t)$ and $\mu^B(\cdot,t)$ turn into each
 other under a time-reversal of the protocol $\lambda(\cdot,s)$.  This occurs most generally when the initial distributions are solely functions of the driving protocol, locally in time.  That is, there exists a function $\phi$ such that $\mu(x,-t) = \phi(\lambda(x,-t))$ and $\mu^B(x,t) = \phi(\lambda(x,t))$.  This includes the cases of equilibrium and nonequilibrium steady states, in which the protocol $\lambda(x,t) \equiv \lambda(x)$ is time-independent.  To be clear, when we speak in this paper of a TFT for entropy production, we mean only the case when local detailed balance and BC1 are satisfied and a protocol-to-distribution mapping $\phi$ exists.

\bigskip

{\bf Remark 3.1.} 
	
\quad In Ref. \cite{Saha}, Saha et. al. employ the techniques of van Zon and Cohen \cite{vanZon2004} to show that for a Brownian particle in a harmonic potential and driven by an arbitrary time-dependent force, the entropy production only satisfies the TFT

\begin{equation}
\label{SAHA}
{\P(\Delta s_{tot}) \over \P(-\Delta s_{tot})} = e^{\Delta s_{tot}}
\end{equation}

 if the particle is initially in thermal equilibrium.  This is indeed a surprising fact because, based on our abstract results and the discussion above, one would {\it never} expect this TFT to be satisfied, but rather the TFT (\ref{GTFT2}), with $Q = \P^-$ and $\varphi = r$.  The reason is that (\ref{SAHA}) fails to distinguish between both the forward and backward path measures and the forward and backward entropy productions, each of which are distinct due to the time-dependent driving.  In fact, as the authors acknowledge, (\ref{SAHA}) holds in the case of equilibrium initial conditions only because of a coincidental relationship between the form of the equilibrium Gibbs measure and the harmonic potential.  Any other potential or initial distribution - even an athermal Gibbs distribution - causes (\ref{SAHA}) to break down.  It is therefore important that no general relationship between the validity of TFTs for entropy production and equilibrium initial conditions be inferred from this study.
 
\quad The situation is similar for another result proved by the authors, that the entropy production of a Langevin system prepared initially in a nonequilibrium state and allowed to relax to equilibrium without external driving does not satisfy the TFT (\ref{SAHA}).  Even in the absence of driving, the path measures $\P$ and $\P^-$ differ in their initial conditions - the nonequilibrium and equilibrium states, respectively - and therefore (\ref{GTFT2}) does not reduce to (\ref{SAHA}).  The entropy production of the system would therefore not be expected to satisfy (\ref{SAHA}).


\subsection{Dissipated Work and Heat}

\quad We now change boundary conditions so that the quantity $S_P$ no longer equals the entropy production of the forward process, but its dissipated work - that is, the entropy that flows from the system to its surrounding reservoir as a result of work having been done on it.  When multiplied by the temperature of the reservoir, this entropy flux equals the difference between the total work $W(-t,t)$ done on the system $+$ environment and the change in Helmholtz free energy of the system.  Our new boundary conditions are 

\begin{equation}
\label{BC2}
\mathrm{(BC2)} \quad \mu(\cdot,-t) = \mu^*(\cdot,-t) \quad \mathrm{and} 
			 \quad \mu^B(\cdot,t) = \mu^*(\cdot,t),
\end{equation}

which, in a laboratory experiment, implies preparing the system initially at equilibrium with respect to the Hamiltonian $H(\cdot,-t)$ and temperature $\beta^{-1}$.  From our previous discussion, BC2 immediately implies that $S_P$ and $S_Q$ represent the same physical quantity in their respective processes.  We further have that the boundary term for $S_P$, replacing $\Delta S$, equals

\begin{eqnarray*} 
\log \frac{\mu^*(X_{-t},-t)}{\mu^*(X_t,t)}
		  & = \beta H(X_t,t) - \beta H(X_{-t},-t) + \log Z(t) - \log Z(-t) \\
		  & = \beta \Delta H - \beta \Delta F,
\end{eqnarray*}

with $\Delta H$ the microscopic energy change and $\Delta F$ the free energy difference between the equilibrium distributions at times $t$ and $-t$.  Adding the current part $\beta Q(-t,t)$, which is justified by local detailed balance and (\ref{HEAT}),

\begin{equation*} \fl
\log d\P/d\P^B = \beta (\Delta H + Q(-t,t)) - \beta \Delta F 
= \beta W(-t,t) - \beta \Delta F,
\end{equation*}

where the RHS is precisely the dissipated work.  

\quad One might naturally assume that if the dissipated work $\beta W(-t,t) - \beta \Delta F$ of a system initially prepared in equilibrium and satisfying both microscopic reversibility and local detailed balance satisfies a TFT, then so would its dissipated {\it heat} $\beta Q(-t,t) = \beta W(-t,t) - \beta \Delta H$.  In fact, it is the absence of the boundary term $\Delta S = \beta \Delta H - \beta \Delta F$ in the latter variable that makes this statement false in general \cite{Seifert,Harris}, particularly in the case of a driven Brownian particle, as was observed by van Zon and Cohen \cite{vanZon2004}.  In our generalized framework, writing $Q(0,t)$ for the heat dissipation of the forward process and $Q^B(0,t)$ for the backward process, the MGF symmetry (\ref{GFT}) with BC1 implies that

\begin{equation*} \fl
\E_P \biggl( e^{\lambda Q(-t,t)} 
	\Bigl( \frac{\mu(X_{-t},-t)}{\mu(X_t,t)} \Bigl)^\lambda \biggl) 
= \E_{P^B} \biggl( e^{-(1+\lambda) Q^B(-t,t)} 
	\Bigl( \frac{\mu(X_t,t)}{\mu(X_{-t},-t)} \Bigl)^{-(1+\lambda)} \biggl)
\end{equation*}

for $-1 \leq \re(\lambda) \leq 0$.  Thus a TFT for heat dissipation holds only if the random variables $\mu(X_{-t},-t)$ and $\mu(X_t,t)$ are equal almost surely.  But even in a nonequilibrium steady state in which the distributions $\mu(\cdot,-t)$ and $\mu(\cdot,t)$ are equal this will not be true, of course, because the microscopic states $X_{-t}$ and $X_t$ will in general differ.  We therefore expect van Zon and Cohen's observation that the heat dissipation of a driven Brownian particle does not satisfy a conventional TFT to hold for the great majority of processes.

\bigskip

{\bf Example 3.2 (Inhomogeneous It\`o diffusions in $\R$$^d$).} 

\quad This generalizes the case of homogeneous diffusions \cite{JiangBook,Maes2000,GeJiang,Qian}, but instead of studying their stationary entropy production as is typically done, we consider their dissipated work due to time-dependent driving.  (For a more in depth, technical exposition on fluctuation theorems for multidimensional diffusions, see Ref. \cite{Chetrite}.)  Let $dX_s = b(X_s,s)ds + \sigma(X_s,s)dB_s$ denote the stochastic differential of the forward process, whose drift vector $b: {\R}^d \times [-t,t] \arrow {\R}^d$ equals minus the gradient of a time-dependent potential $H(x,s)$, to which the system is initially equilibrated at time $-t$, and where $\sigma: {\R}^d \times [-t,t] \arrow M_{d \times m}(\R)$, with $B_s$ an $m$-dimensional Brownian motion.  The only assumptions we make are that $b$ and $\sigma$ are continuous in time and satisfy the usual Lipschitz continuity requirements in space for a weakly unique solution \cite{Oksendal}.

\quad The protocol-reversed differential associated with the forward differential is $dY_s = b(Y_s,-s)ds + \sigma(Y_s,-s)dB_s$, with initial condition $\mu^-(\cdot,-t) = \mu^*(\cdot,t)$ (i.e., BC2).  This in turn yields the backward process $Z_s = Y_{-s}$, whose quadratic variation process $[Z]_s$ is identical to that of the forward process: Letting $\mathcal{P} = \{t_j\}_{0 \leq j \leq n}$ denote a partition of $[-t,s]$,

\begin{eqnarray*}
[Z]_s 
& = \lim_{||\mathcal{P}|| \arrow 0} \sum_{j=0}^n \bigl| Z_{t_{j+1}} - Z_{t_j} \bigl|^2
  = \lim_{||\mathcal{P}|| \arrow 0} \sum_{j=0}^n \bigl| Y_{-t_{j+1}} - Y_{-t_j} \bigl|^2 \\
& = \int_{-s}^t \sigma(\cdot,-u)^2 dB_u = \int_{-t}^s \sigma(\cdot,u)^2 dB_u = [X]_s
\end{eqnarray*}   

The Girsanov theorem therefore guarantees equivalence of $\P$ to a measure $\P'$ which is identical to $\P^B$ except having initial distribution $\mu(\cdot,-t)$.  But $d\P'/d\P^B(\omega)$ then equals $\mu(X_{-t}(\omega),-t)/\mu^*(X_t(\omega),t)$, which is finite by the positivity of Gibbs measures, and so we conclude $\P \sim \P' \sim \P^B$.  This confirms that the dissipative work satisfies a TFT.

\quad Note that the forward path measure will not in general be equivalent to either the protocol-reversed or path-reversed path measures, as it is in the homogeneous diffusion case, because their corresponding quadratic variation processes do not coincide.  This implies that the usual TFT $P(W_d = z)/P(W_d = -z) = e^z$ for dissipated work will not hold.


\subsection{Tautological interpretation of transient fluctuation theorems}

\quad Consider equation (\ref{GTFT2}), which is the distributional form of the TFT identity for virtually all processes of relevance in physics - those whose thermodynamic variables have a discrete or continuous distribution.  With the identifications $P \equiv \P$, $Q \equiv \P^-$ and $\varphi \equiv r$ made in section 3.1, and using Lemma 2.1,

\begin{eqnarray*}
Q(S_Q(\omega) = -x) 
& = \P^- \biggl( \log \frac{d\P^-}{d(r\P)}(\omega) = -x \biggl) \\
& = \P^- \biggl( \log \frac{d(r\P)}{d\P^-}(\omega) = x \biggl) \\
& = \P^- \biggl( \log \frac{d\P}{d\P^B}(r\omega) = x \biggl) \\
& = \P^B \biggl( \log \frac{d\P}{d\P^B}(\omega) = x \biggl). \\
\end{eqnarray*}

implying that (\ref{GTFT2}) can be re-expressed as 

\begin{equation*}
\P \biggl( \log \frac{d\P}{d\P^B)}(\omega) = x \biggl) 
= e^x \, \P^B \biggl( \log \frac{d\P}{d\P^B}(\omega) = x \biggl).
\end{equation*}

\quad Recalling the discussion in section 3.2, this equation states the following tautology:  "Spacetime curves that are $e^x$ times more likely to be realized in a forward-time universe than a backward-time universe [i.e., $\{\omega \in \Omega: \log d\P/d\P^B(\omega) = x\}$] are $e^x$ times more likely to be realized in a forward-time universe than a backward-time universe."  The TFT identity, and hence the thermodynamic TFTs it generalizes, are therefore simply mathematical representations of a self-evident statement.  This is true in particular for the case of homogeneous processes, in which $\P^- = \P$ (modulo boundary terms) and the familiar TFT

\begin{eqnarray*} \fl
\P \biggl( \log \frac{d\P}{d(r\P)}(\omega) = x \biggl) 
& = e^x \, \P \biggl( \log \frac{d\P}{d(r\P)}(\omega) = -x \biggl) \\
& = e^x \, r\P \biggl( \log \frac{d\P}{d(r\P)}(\omega) = x \biggl)
\end{eqnarray*}

has the interpretation that "trajectories that are $e^x$ more likely to be observed than their time-reversals are $e^x$ more likely to be observed than their time-reversals."  

\quad If one is to ascribe nontrivial content to TFTs, therefore, it cannot be to the theorems themselves, which are "obvious", but to the fact that the thermodynamic variable in question can be represented as a logarithmic Radon-Nikodym derivative.  This, after all, is the distinction between the entropy and dissipated work of a process and its dissipated heat, with only the former two satisfying a TFT in general.


\section{Domain of convergence of the MGF and breakdown of asymptotic fluctuation theorems}

\subsection{Definition and breakdown of AFTs}

\quad It was proved in section 2 that the MGF of entropy production in the forward process, $\E_P(e^{\lambda S(0,t)})$, is only guaranteed to converge for $-1 \leq \re(\lambda) \leq 0$, a fact missed by previous studies.  The significance of this fact is that processes for which this function does not converge in a neighborhood of the origin cannot satisfy an AFT.  To see this, recall that the entropy production of a homogeneous process $(X_t)_{t \geq 0}$ (i.e., one that is driven time-independently) satisfies an AFT with speed $\varphi(t)$ when the quantity $S(-t,t)/\varphi(t)$ satisfies a large deviation principle (LDP) with speed $\varphi(t)$ \cite{Touchette}, whose corresponding rate function $I(z)$ satisfies the Gallavotti-Cohen symmetry $I(z) - I(-z) = -z$.  Here $\varphi(t)$ is some monotonically increasing continuous function satisfying $\varphi(t) \arrow \infty$ as $t \arrow \infty$, and $I(z)$ is a nonnegative, lower semi-continuous function such that for all intervals $A \subset \R$,

\begin{equation*}
\lim_{t \arrow \infty} \frac{1}{\varphi(t)} \log P(X_t \in A) = - \inf_{z \in A} I(z).
\end{equation*}

The Gallavotti-Cohen symmetry is the infinite time analogue of (\ref{GTFT2}), which held only for the finite time distribution of entropy production.  Just as the rate function generalizes the finite time distribution of $S(-t,t)$, the free energy 

\begin{equation*}
c(\lambda) = \lim_{t \arrow \infty} \frac{1}{\varphi(t)} \log \E_P 
	\bigl( e^{\lambda S(-t,t)} \bigl)
\end{equation*}

generalizes its MGF.  It is in the relationship between the rate function and free energy that the domain of the MGF becomes relevant.  

\quad By Varadhan's theorem \cite{Varadhan}, the Legendre-Fenchel transform of $I(z)$ yields $c(\lambda)$.  This implies that if the MGF of $S(-t,t)$ does not exist in a neighborhood of the origin for large $t$, then the free energy does not exist there either for {\it any} choice of $\varphi$, meaning that $I(z)$ is not defined in a neighborhood of its minimum $z^*$, which would otherwise be the almost sure limit of $S(-t,t)/\varphi(t)$.  (When $c(\lambda)$ is differentiable at $0$, $z^* = c'(0)$ and the time-averaged entropy production converges exponentially to this value \cite{Shargel}.)  As it is precisely the existence of the rate function in a neighborhood of $z^*$ that guarantees almost sure convergence to this value (i.e., the strong law of large numbers) as well as the distribution of fluctuations about it \cite{Touchette}, this scenario corresponds to a breakdown in the LDP, and hence the AFT.

\quad In fact, even though it has been shown that the heat dissipation $Q(-t,t)$ of a process does not in general satisfy a TFT, the argument above holds for it as well.  In this case, the failure of the MGF of $Q(-t,t)$ to exist in a neighborhood of $\lambda=0$ results in the breakdown of an AFT for the time-averaged heat dissipation and, hence, the time-averaged entropy production (due to the finiteness of the boundary term $\Delta S$), despite the fact that the latter satisfies a TFT over arbitrary timescales.  We consider this situation in the detailed example below.  

\subsection{Example: Strongly biased birth-death chains}

\subsubsection{The model}

\quad To illustrate how the divergence of the MGF for heat dissipation can come about, we consider the example of a continuous-time birth-death chain $X_t$ on the nonnegative integers $j \geq 0$, representing the dynamics of a population.  The chain hops from site $j$ to $j+1$ with rate $p_j$ and left to $j-1$ with rate $q_j$, corresponding to a birth or death in the population, respectively.  For later simplicity, we define the process only on the time interval $[0,\infty)$, with $X_0 \equiv 0$, so that the reversal of the path segment $\omega|_{[0,t]} \in D(0,t)$ is $r(\omega)_s = \lim_{s' \uparrow t-s} \omega_{s'}$, which preserves path right-continuity.  None of our results are affected by defining the process on the halfline instead of all of $\R$.  We further take $p_j + q_j = 1$ so that the mean holding time at every site is $1$ second, and restrict the argument $\lambda$ of the MGF to the real axis, since the free energy, at least as it is employed in large deviation theory, is defined only on the reals.  Note that the former constraint implies that the process makes only a finite number of hops almost surely in a finite time interval, so as long as we take $p_j>0$ for $j \geq 0$ and $q_j>0$ for $j \geq 1$, the forward and backward path measures restricted to that interval will be equivalent.

\quad Following Lebowitz and Spohn and our discussion in section 3, $Q(0,t)$ is incremented by $\log p_j/q_{j+1}$ every time the particle hops right from $j$ and $\log q_j/p_{j-1}$ every time it hops left.  The rate in the denominator refers to the corresponding reversed movement in the backward process.  That $Q(0,t)$ only depends on the final state $X_t$ can be seen by noting that of the $N_t$ hops made until time $t$, exactly $(N_t - X_t)/2$ rightward ones from $j$ to $j+1$ are compensated by leftward ones from $j+1$ to $j$, whose contributions to $Q(0,t)$ cancel.  What remains are contributions made from rightward hops at the first $X_t$ sites, so that 

\begin{equation}
\label{HEAT_DISSIPATION}
Q(0,t) = \log \prod_{j=0}^{X_t - 1} \frac{p_j}{q_{j+1}}.
\end{equation}

The MGF for $Q(0,t)$ can therefore be written as

\begin{eqnarray} 
\fl M_Q(\lambda,t) 
	& \equiv \E_P \bigl( e^{\lambda Q(0,t)} \bigl) \nonumber \\
\fl & = \sum_{n=0}^\infty P(N_t = n) \sum_{k=1}^{n} P(X_t = k | N_t = n)
	\biggl( \, \prod_{j=0}^{k-1} \frac{p_j}{q_{j+1}} \biggl)^\lambda + \, C_0,
\label{MGF}
\end{eqnarray}
 
where $C_0$ represents the contribution from the $k=0$ term in the inner sum, which is independent of $\lambda$ and $t$.  As this constant factor does not affect any of our results for the free energy, which only depends on derivatives and scaled limits of $M_Q(\lambda,t)$, we omit it from future calculations. 
 
\quad We consider two choices for the rates $p_j, q_j$, each of which biases the chain toward the right.  The first choice, $p_j/q_j = \alpha > 1$, we will show, results in a convergent MGF about the origin and a valid AFT for $Q(0,t)$ for all $t \geq 0$.  It is a simple illustration that AFTs can apply to processes that not only do not possess a limiting stationary distribution (whether strict, periodic or quasistationary), but are non-recurrent.  This is also a special case of the AFT proved in Ref. \cite{HarrisTouchette} for non-Markovian simple random walks.  For the stronger bias $p_j/q_j = 2^j$, on the other hand, we show that the MGF fails to exist for $\lambda > 0$ and so no LDP is satisfied for any choice of $\varphi(t)$.  In essence, what fuels the breakdown in this case is that the typical irreversibility of trajectories consisting of arbitrarily large numbers of hops, represented, via (\ref{HEAT_DISSIPATION}), by their heat dissipation, dwarfs their improbability under $P$.  

\subsubsection{Constant Bias}

\quad We begin with the case $p_j/q_j = \alpha > 1$ for $j \geq 1$, with the boundary condition $p_0 = 1$ and $q_0 = 0$.  These rates might correspond to a population of cells whose common division rate is $\alpha$ times greater than their common death rate.  The constraint $p_j + q_j = 1$ corresponds to slowing down the dynamics for larger populations, an effect which does not alter our analysis as all waiting times have canceled in the definition of $Q(0,t)$.  In light of the constraint $p_j + q_j = 1$, we have $p_j = \alpha / (\alpha + 1)$ and $q_j = 1/(\alpha + 1)$, implying by (\ref{HEAT_DISSIPATION}) that $Q(0,t) = \log [\alpha^{X_t - 1}(\alpha + 1)]$ for $X_t \geq 1$ and $Q(0,t)=0$ for $X_t=0$.  

\quad To prove a LDP, we show that the free energy

\begin{equation}
\label{HEAT_FREE_ENERGY}
c_Q(\lambda) = \lim_{t \arrow \infty} \frac{1}{t} \log M_Q(\lambda,t)
\end{equation}

exists and is differentiable, indicating a simple linear speed for the LDP.  We begin by bounding $|M_Q(\lambda,t)| = M_Q(\lambda,t)$ above and below by exponentials in $t$ (with prefactors) to confirm that the speed $\varphi(t)=t$.  Using the fact that $N_t$ is Poisson with intensity $t$, our upper bound is

\begin{eqnarray}
\label{MGF_INITIAL}
M_Q(\lambda,t)
& = \sum_{n=0}^\infty \frac{e^{-t} t^n}{n!} \sum_{k=1}^{n} P(X_t = k | N_t = n)
	\bigl( \alpha^{k-1}(\alpha + 1) \bigl)^{\lambda} \\
& < e^{-t} \sum_{n=0}^\infty \frac{t^n}{(n-1)!} (\alpha + 1)^{\lambda n} \nonumber \\
& = t (\alpha + 1)^\lambda \exp [-t + t(\alpha + 1)^\lambda], \nonumber
\end{eqnarray}

so that 

\begin{equation}
\label{UPPER_BOUND}
\limsup_{t \arrow \infty} \frac{1}{t} \log M_Q(\lambda,t)
	< (\alpha + 1)^\lambda - 1.
\end{equation}

Keeping only the $k=n$ terms in (\ref{MGF_INITIAL}) and noting that

\begin{equation}
\label{NTH_TERM}
P(X_t = n|N_t = n) = \prod_{j=0}^{n-1} \frac{p_j}{p_j + q_j} 
	= \biggl( \frac{\alpha}{\alpha + 1} \biggl)^{n-1},
\end{equation}

we have the lower bound

\begin{eqnarray*}
M_Q(\lambda,t) 
& > \sum_{n=0}^\infty \frac{e^{-t} t^n}{n!} 
	\biggl( \frac{\alpha}{\alpha + 1} \biggl)^{n-1} \alpha^{\lambda n}
  = \frac{\alpha + 1}{\alpha} e^{-t} \sum_{n=0}^\infty \frac{1}{n!}
	\biggl( \frac{t \alpha^{1 + \lambda}}{\alpha+1} \biggl)^n \\
& = \frac{\alpha + 1}{\alpha} 
	\exp \biggl[t \biggl( \frac{\alpha^{1 + \lambda}}{\alpha+1} - 1 \biggl) \biggl],
\end{eqnarray*}

implying that

\begin{equation}
\label{LOWER_BOUND}
\liminf_{t \arrow \infty} \frac{1}{t} \log M_Q(\lambda,t) 
> \frac{\alpha^{1 + \lambda}}{\alpha+1} - 1 \geq -1,
\end{equation}

uniformly in $\lambda$.  

\quad Together, (\ref{UPPER_BOUND}) and (\ref{LOWER_BOUND}) verify that {\it if} the limit (\ref{HEAT_FREE_ENERGY}) exists, the speed $\varphi(t)$ must equal $t$.  We establish this limit by proving that $\frac{1}{t} \log M_Q(\lambda,t)$ is monotonically increasing in $t$ for large $t$, and therefore converges to a finite value, bounded above by the $\limsup$ (\ref{UPPER_BOUND}).  To this end, define $M_Q^1(\lambda,t)$ to be the MGF of $Q(0,t)$ with respect to the process which begins at site $1$ instead of $0$.  It is easy to show that $M_Q(\lambda,t)$ satisfies the backward equation

\begin{equation*} \fl
\frac{\partial}{\partial t} M_Q(\lambda,t) = 
p_0 \, e^{\lambda \log \frac{p_0}{q_1}} M_Q^1(\lambda,t) - M_Q(\lambda,t)
= (\alpha + 1)^{\lambda} M_Q^1(\lambda,t) - M_Q(\lambda,t).
\end{equation*}
 
Since $M_Q(\lambda,t)$ and $M_Q^1(\lambda,t)$ can differ $\forall t$ at most by $(1 + \alpha)^\lambda$, corresponding to the heat released by an immediate jump from $0$ to $1$ at $t=0$, and both grow to infinity as $t \arrow \infty$,

\begin{equation*}
\frac{\partial}{\partial t} \log M_Q(\lambda,t) = (\alpha + 1)^\lambda - 1 + o(1),
\end{equation*}

where the $o(1)$ term vanishes in this limit.  But by the strict inequality in (\ref{UPPER_BOUND}), there exists an $\epsilon > 0$ such that $\log M_Q(\lambda,t) \leq t( (\alpha + 1)^\lambda - 1 - \epsilon)$  for large $t$.  This implies that 

\begin{equation*} \fl
\frac{\partial}{\partial t} \biggl[ \frac{1}{t} \log M_Q(\lambda,t) \biggl]
\geq \frac{1}{t^2} \Bigl( t \bigl( (\alpha + 1)^\lambda - 1 + o(1) \bigl)
	- \, t \bigl( (\alpha + 1)^\lambda - 1 - \epsilon \bigl) \Bigl) 
= \frac{\epsilon + o(1)}{t},
\end{equation*}

which is nonnegative for large $t$ - exactly what was required.

\quad Having established convergence of the free energy $\forall \lambda \in \R$, we now turn to its differentiability.  As $c_Q(\lambda)$ is convex and differential operators commute with limits of convex functions, upon taking a derivative of (\ref{HEAT_FREE_ENERGY}), we may pass the operator through the limit and ultimately to the MGF inside.  The derivative of this infinite sum is then evaluated term-wise, which is justified once it is clear that the resulting sum converges absolutely and uniformly on compact sets:

\begin{eqnarray*}
\fl \Bigl| \frac{\partial}{\partial \lambda} M_Q(\lambda,t) \Bigl|
& = \biggl| e^{-t} \sum_{n=0}^\infty \frac{t^n}{n!} 
	\sum_{k=1}^{n} P(X_t = k | N_t = n) \bigl( \alpha^{k-1}(\alpha + 1) 
	\bigl)^{\lambda} \, \log \bigl( \alpha^{k-1}(\alpha + 1) 
	\bigl) \biggl| \\
\fl & \leq e^{-t} \sum_{n=0}^\infty 
	\frac{ \bigl[ t (\alpha + 1)^\lambda \bigl]^n}{(n-1)!} \cdot n \log (\alpha + 1)
\end{eqnarray*}

Local uniform convergence clearly holds (by the Ratio Test, for instance), and so we may conclude that the free energy $c_Q(\lambda)$ is differentiable.  

\subsubsection{Strong Bias}

\quad Having proved a LDP and associated AFT for the constant bias case, we now show that for the exponentially biased rates $p_j/q_j = 2^j$, no LDP is possible for any choice of speed $\varphi(t)$.  This is a somewhat surprising result, since one might suspect that no matter how fast heat is dissipated by a process, there exists a time scaling under which the distribution of its fluctuations has a weak limit, analogous to a central limit theorem.  Interestingly, the MGF exists and a LDP is satisfied even for the linearly increasing bias $p_j/q_j = j$, although there is no simple representation of the LDP speed for these rates.  

\quad In the present case, the rates can be solved as $p_j = 2^j/(2^j + 1)$ and $q_j = 1/(2^j + 1)$ (again, with $p_0 = 1$ and $q_0 = 0$), whose associated heat dissipation by time $t$ is

\begin{equation*}
Q(0,t) = \log \, (2^{X_t} + 1) \prod_{j=1}^{X_t - 1} 2^j. 
\end{equation*}

Keeping only the $k=n$ term in its definition (\ref{MGF}) and recalling the first equality in (\ref{NTH_TERM}), we obtain the following inequality for $M_Q(\lambda,t)$:

\begin{equation}
\label{MGF_STRONG_BIAS}
\eqalign{
M_Q(\lambda,t) 
& = \sum_{n=0}^\infty \frac{e^{-t} t^n}{n!} \sum_{k=1}^{n} 
	P(X_t = k | N_t = n) 
	\biggl( (2^k + 1) \prod_{j=1}^{k - 1} 2^j \biggl)^\lambda \\
& \geq \sum_{n=0}^\infty e^{-t} t^n \biggl( \prod_{k=1}^n 
	\frac{2^k}{2^k + 1} \biggl)
	\prod_{j=1}^n \frac{2^{\lambda j}}{j}}
\end{equation}

Defining $\eta \equiv \prod_{k=0}^\infty \frac{2^k}{2^k + 1} > 0$,  

\begin{eqnarray*} 
|\log \eta|
 = \biggl| \sum_{k=0}^\infty \log \biggl( 1 - \frac{1}{2^k + 1} \biggl) \biggl|
 = \biggl| \sum_{k=0}^\infty \biggl[ - \frac{1}{2^k + 1} 
	+ \frac{C}{2!} \frac{1}{(2^k + 1)^2} \biggl] \biggl| < \infty,
\end{eqnarray*}

where $C$ is the appropriate coefficient in the exact $2^{\mathrm{nd}}$-order Taylor expansion of $\log(x)$.  We therefore see that $\eta$ is finite, and so $M_Q(\lambda,t) \geq e^{-t} \sum_{n=0}^\infty t^n a_n(\lambda)$, where the sequence $a_n(\lambda) \sim \eta \prod_{j=1}^n 2^{\lambda j} / j$ tends to infinity $\forall \, \lambda > 0$.  Therefore $M_Q(\lambda,t)$ diverges $\forall t>0$ for $\lambda > 0$, and we are done.

\quad As we remarked earlier and can be seen in (\ref{MGF_STRONG_BIAS}), the divergence of the MGF in the strong bias case comes from the domination of the exponential heat dissipation term $\prod_{j=1}^n 2^{\lambda j}$ over the $1/n!$ term that weights the probability of a trajectory with $n$ hops by time $t$.  Physically, this means that the typical heat dissipation associated with the tail event in which the system hops a large number of times over a finite time interval dwarfs the improbability of that event, leading to the divergence of all of its moments.

\section{Conclusion}

\quad The abstract results of this paper imply several consequences for the interpretation of TFTs.  First and foremost, the mathematical identity underlying TFTs, represented in its distributional form by the fluctuation symmetry (\ref{GTFT2}), is a very general one.  As demonstrated in Theorems 2.2 and 2.3, it holds for logarithmic Radon-Nikodym derivatives between processes that need have no relation to each other than to put positive probability on the same set of paths - hence the example in the introduction.  In particular, the fluctuation symmetry does not require the processes to be related by a protocol, trajectory or field reversal, or any other self-inverse "reversal" transformation employed in physics derivations of TFTs, in order to hold.  

\quad The generality of the fluctuation symmetry can even be taken a step further.  While the fluctuation symmetry does not hold in general for the heat dissipated by a process, which cannot be represented as a logarithmic derivative, in the case of entropy production and dissipated work the symmetry merely expresses a self-evident statement in terms of the arrow of time.  If one is to ascribe meaning to TFTs for these quantities, and therefore to the second law of thermodynamics which they imply as a consequence, it must be to the representation of the two thermodynamic variables as logarithmic derivatives.  This puts a very different face on Loschmidt's paradox, which has always been cast as a dynamics problem, rather than one of mathematical representation.

\quad Another result that comes out of Theorem 2.2 is that the MGFs of the thermodynamic variables appearing in the MGF form of the identity are not guaranteed to exist in an open neighborhood of the origin.  As the detailed example of strongly biased birth-death chains demonstrated, this can lead to a breakdown in the LDP of the thermodynamic variable in question under {\it every} possible time-scaling, and hence a breakdown of the associated AFT.  The implication is that the fluctuations of a variable may satisfy a TFT over arbitrarily timescales, but the fluctuations of its time-average would not satisfy an AFT.  

\ack{The author would like to thank Tom Chou and Thomas Richthammer for helpful discussions.  This work was supported by grants from the NSF (DMS-0349195) and the NIH (K25 AI41935), as well as the VIGRE Graduate Fellowship.}

\bigskip

{}

\end{document}